\documentclass[journal]{IEEEtran}
\usepackage{amsmath,amsfonts}
\usepackage{array}
\usepackage[table]{xcolor}
\usepackage{graphicx}
\usepackage{cite}
\usepackage[hidelinks]{hyperref}
\usepackage{orcidlink}
\hypersetup{
	pdftitle={Goal-Oriented Communications for Interplanetary and Non-Terrestrial Networks},
	pdfauthor={Elif Uysal and Aimin Li},
	pdfsubject={Goal-oriented networking for SpaceNets},
	pdfkeywords={6G NTN, age of information, delay and disruption tolerant networking, random access, satellite IoT, space relay, Bundle Protocol}
}

\definecolor{SpaceTableHead}{HTML}{2F5D8C}
\definecolor{SpaceTableBand}{HTML}{F3F7FB}
\definecolor{SpaceTableRule}{HTML}{B8C6D9}
\definecolor{PillarSampling}{HTML}{0B6FA4}
\definecolor{PillarAccess}{HTML}{B35A12}
\definecolor{PillarFlow}{HTML}{7030A0}
\makeatletter
\def\section{\@startsection{section}{1}{\z@}{.85ex plus .25ex minus .15ex}{.35ex plus .1ex}{\normalfont\large\bfseries}}
\def\subsection{\@startsection{subsection}{2}{\z@}{.55ex plus .2ex minus .1ex}{.2ex plus .08ex}{\normalfont\normalsize\itshape}}
\let\oldthebibliography\thebibliography
\let\endoldthebibliography\endthebibliography
\renewenvironment{thebibliography}[1]{%
	\oldthebibliography{#1}%
	\setlength{\itemsep}{0pt}%
	\setlength{\parskip}{0pt}%
	\setlength{\parsep}{0pt}%
}{\endoldthebibliography}
\makeatother

\begin{document}
	\title{Goal-Oriented Communications for Interplanetary and Non-Terrestrial Networks}
	
	\author{Elif Uysal~\orcidlink{0000-0002-7258-4872},~\IEEEmembership{Fellow,~IEEE} and 
		Aimin Li~\orcidlink{0000-0003-3687-4378},~\IEEEmembership{Member,~IEEE}
		\vspace{-2.5em}
		\thanks{The authors are with the \href{https://cng-eee.metu.edu.tr/}{Communication Networks Research Group}, Electrical and Electronics Engineering Dept., METU, Ankara, Turkey. The work is supported by the European Union (through ERC Advanced Grant 101122990
			GO SPACE-ERC-2023-AdG.)}
	}

\maketitle

\begin{abstract}
	The recent surge in satellite connectivity and deep space exploration calls for scalable communication network architectures that can effectively support increasing numbers of bursty flows, such as those occurring in remote monitoring and {in-orbit computation}. Yet communications over space links face challenges beyond delay alone: a contact window is a rare chance, and it can be wasted by data that loses task value before feedback returns. This article presents a \textit{goal-oriented networking} framework for SpaceNets, organized around a common question: \textit{will the information still matter when it arrives?} The framework has three pillars: (1) goal-oriented sampling and scheduling that can handle highly variable delay processes with memory, (2) grant-free access policies replacing exogenous arrivals with goal-oriented traffic shaping, and (3) flow-control and routing mechanisms that rank, reprocess, forward, or discard packets/bundles according to goal-oriented metrics at intermediate relays. Together, these mechanisms form a \textit{store-compute-forward} perspective in which application goals guide protocol-level decisions across sampling, access, and relay operation.
\end{abstract}


\section{Introduction}
\IEEEPARstart{R}{esponding} to the demand for a scale-up of the Internet of Things (IoT), new performance optimization tools for Machine Type Communications (MTC) such as Age of Information (AoI), Value of Information, and variants, have been developed within the last decade.
\textit{Goal-oriented networking} exploits these new tools to advance beyond merely ensuring high-throughput and reliable data transmission; instead, it aims to maximize the relevance or utility of transported data with respect to the intended task objectives at the moment the data arrives at its destination ~\cite{uysal2021semantic}.
	
	Concurrently with these research innovations related to MTC, both industry and governments have been investing in a connectivity infrastructure in Earth Orbit, in the Cislunar Space, and beyond: The LEO orbit is already crowded with large commercial constellations. Non-terrestrial Networks (NTN), which use base stations in orbit, are poised to be a major part of 6G networks. In the deep-space realm, a Solar System Internet~\cite{SWGReport} and communication standards to support it are being developed. This interplanetary Internet is envisioned as a heterogeneous network of nodes, ranging from end-of-mission spacecraft repurposed as routers to rapidly growing fleets of CubeSats and sensors deployed on planetary surfaces.

	Future SpaceNets will increasingly support bursty, low-duty-cycle traffic generated by connected space assets for remote inference, decision making, learning, and automation, in addition to the traditional bulk science data. Representative scenarios include training digital twins from data streams collected by Martian rovers or lunar robotic construction systems, delivering time-critical space-weather alerts from an L1 coronagraph telescope to the International Space Station, and tracking maritime cargo through intermittently accessible non-geostationary satellite constellations. These examples share the same pressure: each message competes for a limited \textit{contact opportunity}, and stale or otherwise \textit{task-irrelevant} data can crowd out queues at routers and relays. The resulting design problem is not only contact-aware data selection, but also contact-aware forwarding/scheduling/prioritization computation: SpaceNets must decide when to generate and transmit the next packet, which packets should be prioritized, reprocessed, forwarded, or discarded across surface nodes, orbital relays, and space routers. This is particularly challenging when the timeliness requirements of these applications range from milliseconds to minutes, while links are delay/disruption-prone, relays are intermittently available, and interplanetary contact windows are limited. In this paper, we present a vision for tackling these {fundamental} challenges, to sustain deep-space ventures and the growing satellite infrastructure.

	\section{SpaceNets Beyond the Bit-Pipe}
	
	Traditionally, interplanetary and satellite communication links have relied on relatively conventional infrastructure that primarily serves as a {\it bit-pipe}~\cite{burleigh2019connectivity}. This model has been adequate for supporting individual missions characterized by bulk transfers of images and video, telemetry/telecommand signals, and proprietary GEO/MEO/LEO satellite systems. However, it does not scale to the rapidly increasing number of space assets and machine-type communication (MTC) flows sending short, time-sensitive packets within shared limited bandwidth and contact windows, often under power constraints. This calls for new ways of using communication resources in SpaceNets.
	
	For example, consider satellite-based weather prediction. The inefficiency of transmitting gigabytes of raw imagery to the ground for subsequent decision-making is evident, given that a significant fraction ($\sim 30\%$) of these images is typically cloud-obscured. {In-orbit goal-aware sampling and scheduling offers a direct remedy. As a second example, consider a deep-space relay.} The absence of persistent connectivity makes it difficult to predict the traffic arriving at the next contact. This renders relay queues prone to overflow, which can be managed only through judicious packet discarding. Such discarding requires prioritization based on metrics that capture the significance of the data, and ideally, how its significance decays with staleness, a relationship that is application-dependent.

	Recent works have proposed mechanisms for
		semantic access, feature scheduling~\cite{gao2024semantic}, and communication-computation
		co-design~\cite{triphaticarlone,QuekQuacking24}. These studies commonly assume persistent connectivity and feedback availability, conditions that are not guaranteed in SpaceNets. In LEO access or deep space scenarios alike, a contact window is a scheduled, finite-duration event: once it closes, no retransmission or feedback is possible. This article therefore studies these lesser-studied goal-oriented communication problems that carry great significance in the SpaceNet environment: \textit{When should sources generate fresh data updates in anticipation of a transmission opportunity?}
		\textit{Which sources should contend during a contact window?} \textit{With respect to the operation of relays and routers: which
			stored packets/bundles to discard or reprocess, what to forward, and through which available next-hop path or contact opportunity?} All of the above questions remain open even with ideal semantic transceivers at
		every hop. 
	
	Answering these questions requires tractable performance metrics for prioritization that can handle scenarios where communication delays exceed the time scale of the underlying system dynamics. Particularly motivated by the need for these {solutions} in the developing NTN and space communication standards, we organize
	goal-oriented networking for space around the following three pillars (illustrated in Fig. \ref{fig:sampling}):
	
	\begin{itemize}
		\item \textcolor{PillarSampling}{\textbf{Pillar~I.}} \textbf{Goal-oriented sampling and scheduling} for inference
		and decision making over delay- and disruption-prone networks;
		\item \textcolor{PillarAccess}{\textbf{Pillar~II.}} \textbf{Goal-oriented random access} (GORA), replacing exogenous
		arrivals with goal-aware traffic shaping and access control;
		\item \textcolor{PillarFlow}{\textbf{Pillar~III.}} \textbf{Goal-oriented flow control and routing} for load uncertainty at contact, long delay, irregular feedback, and storage constraints.
	\end{itemize}
	The next section provides a structured overview of the SpaceNet environment and
		relates its key challenges to the corresponding pillars of the proposed
		goal-oriented networking framework. 
	
	\section{Challenges in SpaceNets and Goal-Oriented Networking}\label{sec:spacenets}
	The classical architecture of SpaceNets, dating back to 1977 when Voyager I and II were launched, has historically supported network configurations with a limited number of space assets. Shortly after NASA deployed the rover Sojourner on Mars in 1997, the Interplanetary Networking Special Interest Group (IPNSIG) started outlining a network infrastructure to eventually support the Solar System Internetwork (SSI) architecture: a complex network of networks that includes landers, probes, orbiters, and rovers~\cite{SWGReport}. This architecture, documented by the Internet Engineering Task Force (IETF) and adopted by the Consultative Committee for Space Data Systems (CCSDS), includes protocols such as Unified Space Link Protocol (USLP), Contact Graph Routing (CGR), Licklider Transmission Protocol (LTP), and the Bundle Protocol (BP), developed specifically for the DTN (Delay and Disruption Tolerant Networks) environment.
	
\begin{figure*}[!t]
		\centering
		\includegraphics[width=.7\textwidth]{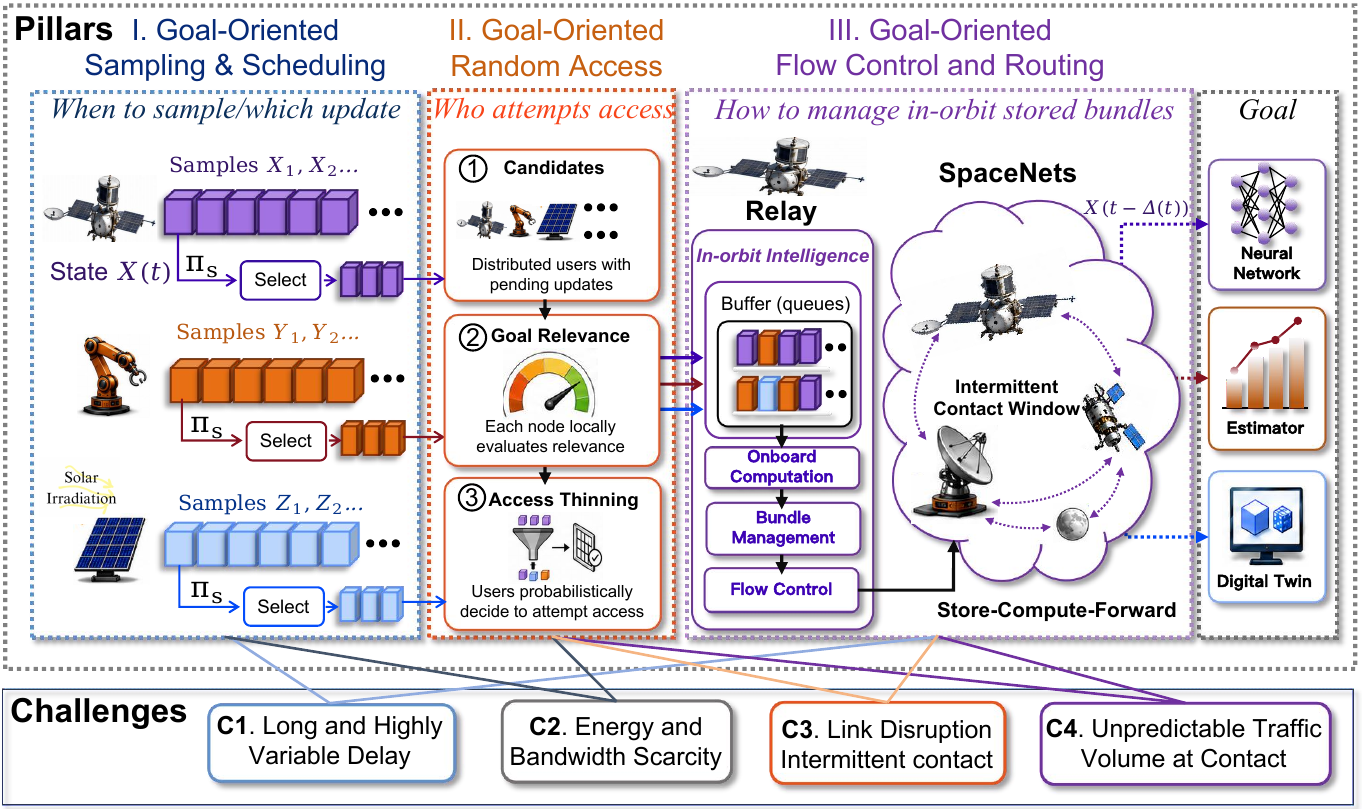}
		\caption{{Pillars of goal-oriented networking vs challenges in SpaceNets. Large and highly variable delay, scarce energy and bandwidth, intermittent contact, and unpredictable traffic volume create irreversible opportunity losses under classical \textit{bit-pipe} operation. Goal-oriented sampling and scheduling, goal-oriented random access, goal-oriented flow control and routing each respond to one or more of these challenges.}}
		\label{fig:sampling}
	\end{figure*}

	While disruption tolerance remains important, another emerging issue is that SpaceNets are expanding scale and traffic diversity~\cite{burleigh2019connectivity}. Beyond traditional bulk science transfers and telemetry/telecommand traffic, they must increasingly support bursty machine-type flows for remote inference, monitoring, automation, and digital-twin maintenance. These flows differ from classical file-transfer traffic because their value is often time- and task-dependent. A measurement that arrives too late, a contact window spent on low-value bundles, or a buffer occupied by stale data may consume an opportunity that cannot be recovered by retransmission. We categorize below four challenges in SpaceNets. Fig.~\ref{fig:sampling} depicts how the challenges are addressed by the three pillars of our vision.
	
	\textbf{Challenge 1 (C1): Large and variable delays.} Long RTTs on interplanetary links often rule out
	standard protocol components such as ARQ (Automatic Repeat Request), \texttt{ping} 
	and Transport Control Protocol (TCP). For example, the Earth--Mars RTT varies between 8.6-42 minutes across the orbital cycle. In such regimes, communication delay can exceed the time scale of the monitored process, so inference and prediction must often rely on data that would be considered stale by terrestrial standards. A source that samples and transmits without accounting for delivery-time value can systematically occupy scarce contact capacity with updates whose relevance has expired by the time they arrive; once the contact window is consumed, no retransmission can recover the lost opportunity. This motivates \textcolor{PillarSampling}{\textbf{Pillar~I}} for delay-aware generation and service, and \textcolor{PillarFlow}{\textbf{Pillar~III}} when relay queueing further compounds propagation delay. 
	
	\textbf{Challenge 2 (C2): Energy and bandwidth scarcity under growing traffic diversity.} SpaceNet traffic now ranges from delay-tolerant science data generated by deep-space probes and telescopes to high-rate mission data and fast time-scale telemetry for monitoring, inference, automation, and digital twins. At the same time, both bandwidth and energy remain severely constrained. Photovoltaic generation is a major power source for spacecraft transceivers, yet available solar power decreases quadratically with distance from the Sun; for example, Saturn, roughly nine times farther from the Sun than Earth, receives only about 1\% of the solar irradiance available
	in Earth orbit. In terrestrial networks, redundant transmission, oversampling, or retransmission may be acceptable fallbacks. In resource-limited SpaceNets, a wasted transmission may consume the energy or downlink capacity needed for a later critical measurement. This motivates \textcolor{PillarSampling}{\textbf{Pillar~I}} for selective information generation and \textcolor{PillarAccess}{\textbf{Pillar~II}} for goal-aware access thinning. 
	
	\textbf{Challenge 3 (C3): Link disruption and intermittent contact.} Routing in Space must respect contact windows that are largely determined by orbital mechanics. For example, CGR plans transmissions by anticipating future contacts and node positions. In multihop settings, CGR requires storage at intermediate nodes while waiting for contact, resulting in a \textit{store-and-forward} structure. Link disruptions may also arise from atmospheric and solar disturbances, such as the impulsive phase of a solar flare, as well as from intermittent energy availability. When a contact opens, forwarding and access decisions must prioritize traffic that most deserve that opportunity; a missed contact may not return until the next orbital period. This motivates \textcolor{PillarAccess}{\textbf{Pillar~II}} for sparse-contact access and \textcolor{PillarFlow}{\textbf{Pillar~III}} for store-and-forward relays. 
	
	\textbf{Challenge 4 (C4): Unpredictable traffic volume at contact.} Even when future contacts are known, the amount of traffic arriving at a relay before the next contact can be difficult to predict. Neighboring nodes may generate traffic according to local events, sensing outcomes, energy availability, or intermittent upstream contacts, while long RTTs make timely feedback unavailable. As feedback may arrive only after a large number of packets are already in flight, reactive congestion control is insufficient. So, the network needs to proactively and autonomously rank, admit, forward, retain or discard packets/bundles at source nodes as well as intermediate contact bottlenecks such as routers and relays, making this the primary operating regime of \textcolor{PillarFlow}{\textbf{Pillar~III}}.
	
		We refer to this operating regime as
		\emph{store-compute-forward}: a goal-oriented generalization of classical
		store-and-forward in which intermediate nodes do not simply preserve queue
		order, but compute goal relevance under delay, storage, energy, and contact
		constraints. In this view, \textcolor{PillarSampling}{\textbf{Pillar~I}}
		defines how information value evolves with generation and delivery time,
		\textcolor{PillarAccess}{\textbf{Pillar~II}} shapes which traffic reaches
		the network, and \textcolor{PillarFlow}{\textbf{Pillar~III}} implements
		prioritization, forwarding, and dropping by computing the remaining value of the packets at a forwarding node. The interplay of the
		three pillars under the store-compute-forward regime is illustrated in
		Fig.~\ref{fig:sampling}, while Table~\ref{tab:use_cases} provides
		representative use-case instantiations.

	\begin{table*}[!t]
		\caption{Representative SpaceNet use cases and the mapping from the
				challenges of Section~\ref{sec:spacenets} (C1--C4) to goal-oriented
				mechanisms. }
		\label{tab:use_cases}
		\centering
		\scriptsize
		\setlength{\tabcolsep}{2.5pt}
		\renewcommand{\arraystretch}{1.04}
		\arrayrulecolor{SpaceTableRule}
		\begin{tabular}{@{}>{\raggedright\arraybackslash}p{0.16\textwidth}
				>{\raggedright\arraybackslash}p{0.16\textwidth}
				>{\raggedright\arraybackslash}p{0.26\textwidth}
				>{\raggedright\arraybackslash}p{0.38\textwidth}@{}}
			\hline
			\rowcolor{SpaceTableHead}
			\textcolor{white}{\textbf{Use case}} &
			\textcolor{white}{\textbf{Goal metric}} &
			\textcolor{white}{\textbf{Operating context (challenges)}} &
			\textcolor{white}{\textbf{Goal-oriented mechanism}} \\
			\hline
			
			\rowcolor{SpaceTableBand}
			Satellite weather observation &
			Forecast-error reduction &
			LEO downlink windows; scarce capacity (C1, C2) &
			\textcolor{PillarSampling}{\textbf{Pillar~I:}}
			select observations with highest downlink forecast value \\
			
			Digital-twin maintenance for spacecraft &
			State-estimation error &
			Multi-hop relays; variable queueing delay (C1, C4) &
			\textcolor{PillarSampling}{\textbf{Pillar~I}} +
			\textcolor{PillarFlow}{\textbf{Pillar~III:}}
			sample by synchronization loss; rank bundles by residual model value \\
			
			\rowcolor{SpaceTableBand}
			Mars rover / lunar robot monitoring &
			Control/localization error &
			Long RTTs; intermittent orbiter contacts (C1, C3) &
			\textcolor{PillarSampling}{\textbf{Pillar~I}} +
			\textcolor{PillarFlow}{\textbf{Pillar~III:}}
			sample by delivery-time value; forward/drop bundles hop-by-hop \\
			
			Lunar surface sensor network &
			Event-detection probability &
			Sparse orbiter passes; many low-duty cycle sensors (C2, C3) &
			\textcolor{PillarSampling}{\textbf{Pillar~I}} +
			\textcolor{PillarAccess}{\textbf{Pillar~II:}}
			couple sampling and access so high-value events contend first \\
			
			\rowcolor{SpaceTableBand}
			Satellite IoT / maritime tracking &
			Tracking error &
			Massive number of users; intermittent visibility; memory storage limitation (C2, C3, C4) &
			\textcolor{PillarSampling}{\textbf{Pillar~I}} +
			\textcolor{PillarAccess}{\textbf{Pillar~II}} +
			\textcolor{PillarFlow}{\textbf{Pillar~III:}}
			goal-aware sampling, access thinning, and bundle ranking \\
			
			Space-weather alerting &
			Deadline-hit rate &
			Rare urgent bursts; strict latency (C1, C2, C4) &
			\textcolor{PillarSampling}{\textbf{Pillar~I}} +
			\textcolor{PillarAccess}{\textbf{Pillar~II}} +
			\textcolor{PillarFlow}{\textbf{Pillar~III:}}
			event-driven sampling, urgent access, and Lifetime-based expiry \\
			
			\hline
		\end{tabular}
		\arrayrulecolor{black}
	\end{table*}

	\section{Pillar I: Goal-Oriented Sampling and Scheduling} 
	\label{sec:sampling}
	The goal-oriented optimization of when information should be generated (sampling), if and when it should be admitted into the network, and which update should be served when communication opportunities are limited (scheduling) address particularly the challenges C1 and C2 as depicted in Fig.~\ref{fig:sampling}. Consider remote monitoring tasks involving, for example, the Dragonfly
	lander on Titan, a construction robot on the Moon, or the solar panels
	of a space probe. Measurements produced at the source are delivered to
	a remote destination, where they support a specific task such as
	inferring the current location of the lander, estimating the position
	of a robot arm, or maintaining the battery state of a spacecraft
	digital twin. These examples correspond to the Mars rover, lunar robot
		monitoring, and digital-twin maintenance scenarios of
		Table~\ref{tab:use_cases}, where the respective goal metrics are
		control and localization error and state-estimation error. In each case, the goal of an update is to reduce an application-level loss, such as estimation error, localization error, or control penalty. In canonical remote-estimation models, this leads to sampling rules that generate a new sample only when the process has changed enough relative to the delay statistics~\cite{li2025optimal}, which can be learned from occasional feedback.
	
	A useful way to gauge the impact of time-varying delay is to express the task performance loss as a function of the \emph{staleness of the data} at the destination. Staleness, represented by the Age of Information, has two causes: the time elapsed between generating fresh measurement samples, and the delay experienced by transmitted samples inside the network. In SpaceNets, the latter may include waiting at interfaces, relays, and routers; samples injected without regard to such delays become \textit{exogenous} arrivals that can occupy queues until they are stale or obsolete. Goal-oriented sampling avoids this failure mode by preventing updates that would \textit{lose relevance by the time they are delivered} from entering the network in the first place. Such adaptation can reduce the error in monitoring severalfold~\cite{li2025optimal}.

	\subsection{Utilizing Age in Goal-Oriented Sampling and Scheduling}
	Consider a flow of packets sent by a source to a remote destination
	over a space connection. Let $u(t)$ be the generation time of the
	newest packet of that flow received by the destination by time~$t$.
	The \textit{age} of this flow at time $t$ is $\Delta(t) = t - u(t)$. Queuing analysis reveals that classical design principles, such as
	the zero-waiting policy that maximizes throughput or minimizes delay,
	can drastically fail to optimize age and functions of age.
	
	In some applications, the distortion or performance loss can be expressed directly in terms of the age of status updates. Consider autonomous rovers exchanging location information with a
		lander or with one another, the Mars rover / lunar robot monitoring
		scenario of Table~\ref{tab:use_cases}, as studied in the multi-agent
		setting of~\cite{triphaticarlone}. Suppose vehicle $1$ tracks the last received location of vehicle $2$, $x_2(t)$, and the age of this information, $\Delta_2(t)$. From the perspective of vehicle~$1$, the radius of the uncertainty region around $x_2(t)$ is the product of the speed of vehicle~$2$ and the age of the last received position information, namely $v_2\Delta_2(t)$. Age-based metrics have also been generalized to account for the usefulness of the next update to the computation at the destination, including Age of Incorrect Information (AoII) and Query AoII.
	
	However, goal relevance is not determined by age alone. In many networked monitoring, robotics, and control systems, not every update is equally informative, and newer updates are not necessarily more valuable. This motivates a joint sampling-and-scheduling formulation: a sampling policy $\pi_s$ decides when samples are generated, while a transmission policy $\pi_t$ decides which queued sample is served. A representative objective is to minimize a long-term average task loss $\mathcal{L}(X(t-\Delta(t)),\hat{X}(t))$, where $\hat{X}(t)$ is the destination-side estimate or decision variable. The loss may be MSE, a control penalty, a localization error, or the cost of missing a target. Note that the delay values experienced by successive transmissions may be dependent. Such delay memory is common in SpaceNets because disruptions, contact windows, and relay queues evolve over time rather than reset independently after each packet.
	
	When the loss function $\mathcal{L}$ is non-monotone in age, as can occur in networked robotics, automation, and inference scenarios, the optimal transmission policy over a delayed channel may choose an older packet from the buffer rather than the freshest one~\cite{shisher2022does}. Hence goal-oriented scheduling in SpaceNets requires {\emph{sampling and scheduling policies that go beyond minimizing age}}. The relevant notion of urgency is not simply ``\textit{freshest first},'' but the expected contribution to task performance at the time of delivery.
	\subsection{Energy and Computational Constraints}
	Space communications can be severely energy constrained, so scheduling policies must respect energy constraints. For example, age-optimal transmission scheduling under energy constraints is achieved by threshold policies where the sender should transmit once the age exceeds a certain threshold value~\cite{BacinogluOhioState}. The threshold will depend on the amount of energy stored, and is in most cases non-increasing in the storage capacity. In other words, the transmitter should hold off on sending the next sample until the previous sample that was sent gets older than a certain duration, which gets longer as the battery level drops. Note that this action is in contrast with what a classical transmission policy trying to maximize throughput would do, and can significantly cut down the communication bandwidth and battery capacity needed (therefore the size and weight) for energy-harvesting space probes and autonomous systems on planet surfaces. 
	
	A practical concern on resource-constrained spacecraft is the onboard cost of goal-oriented policies. Across several goal-oriented sampling formulations, including energy-constrained scheduling~\cite{BacinogluOhioState} and delay-aware sampling~\cite{li2025optimal}, the resulting policies often exhibit low-dimensional or threshold-type structures: the source acts when a locally tracked quantity, \textit{e.g.}, age, estimated penalty reduction,	or stored energy, exceeds a precomputed level. This reduces the per-flow online decision to a simple comparison operation per epoch, requiring no real-time optimization. More expensive optimization or learning can be performed offline, or refreshed from the ground at contact-cycle time scales. This separation between slow statistical learning and fast local decisions makes goal-oriented sampling compatible with the computational constraints of onboard hardware. 
	
	\vspace{0.4em}
	\noindent\fbox{
		\parbox{.95\columnwidth}{
			\textbf{Takeaway}. Goal-oriented sampling and scheduling turns \textit{priority/urgency} from heuristic labels into a delivery-time test: An update warrants transmission only
			if its expected delivery-time value justifies the contact,
			energy, and queueing resources it consumes.}}
	
	\section{{Pillar II: Goal-Oriented Random Access}}
	GORA addresses the access-layer decisions arising from energy and bandwidth scarcity (C2), link disruption and intermittent contact (C3), and distributed contention among many low-duty-cycle sources, as depicted in Fig. \ref{fig:sampling}. When many nodes share a sparse access opportunity, the central question becomes \emph{which sources should attempt access} when a contact becomes available. 
	
\begin{figure}[!t]
		\centering
		\includegraphics[width=.43\textwidth]{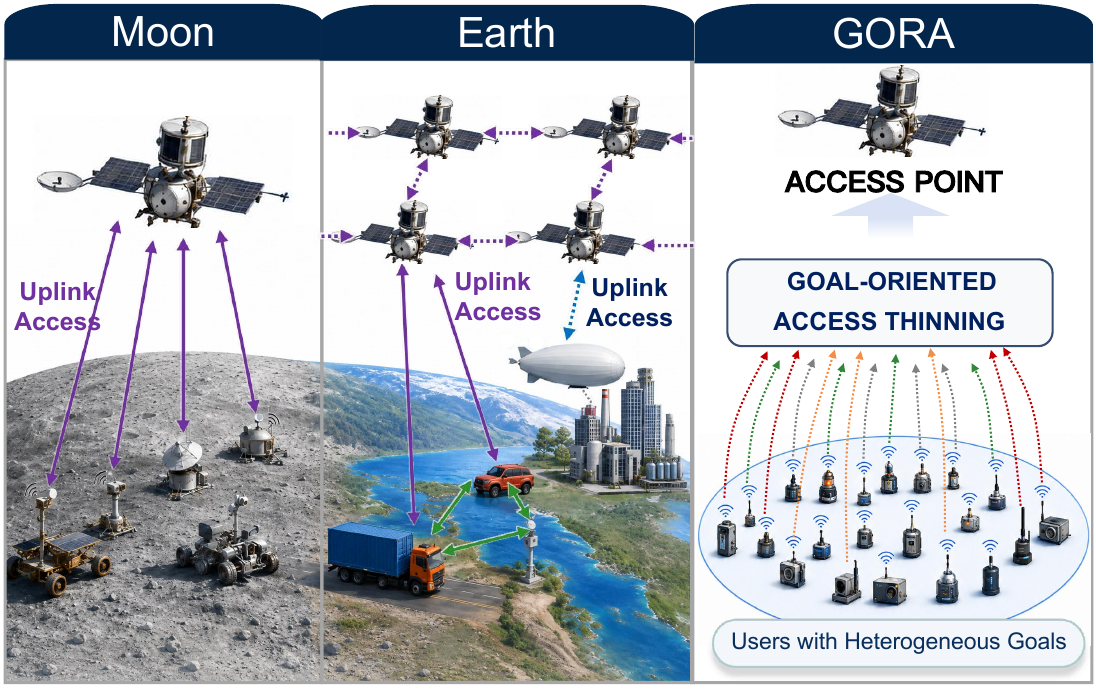}
		\caption{{Random access scenarios on the Lunar surface (landers and sensors to an orbiter), and on Earth (sensors reaching a ground relay or directly reaching a  base station on satellite/High Altitude Platform, possibly using nB-IoT or LPWAN protocols.}}
		\label{fig:randomaccess}
	\end{figure}

	Random-access protocols are particularly relevant for NTN and satellite IoT, where low-duty-cycle users generate short packets and demand-assignment access can be inefficient~\cite{satelliteSurvey21}. A satellite beam may cover a large area, so a single satellite access point may serve massive numbers of IoT nodes. Similar contention arises in lunar and deep-space scenarios, where distributed sensors must reach a lander, an orbiter, or a relay, as illustrated in Fig.~\ref{fig:randomaccess}.
	
	\subsection{From Throughput to Goal Relevance}
	
	ALOHA-type random access remains practical in NTN and satellite IoT, including NB-IoT-over-NTN \cite{3gpp36763} and LPWAN-based systems~\cite{satelliteSurvey21}. These protocols, however, suffer from a well-known throughput limitation. \textit{Coded random access} methods have therefore been proposed to increase network capacity by resolving packet collisions through advanced decoding and successive interference cancellation (SIC), relying on cross-layer PHY/MAC operation at the receiver. For example, the CRMA protocol~\cite{MunariModernRandomAccess} can approximately double the throughput of ordinary slotted ALOHA, whose throughput is limited to about $0.37$ packets per slot. Coded random access has also been adopted in practical satellite systems, such as DVB-RCS2, supporting its feasibility in space communications.
	
	Coded random access and goal-oriented access are complementary. The former resolves collisions at the receiver, whereas the latter reduces contention at the source by activating only users whose data has sufficient goal value. Thus, GORA does not replace advanced random-access receivers; it provides a source-side traffic-shaping layer that can be combined with them.
	
	Yet, \emph{high throughput alone does not guarantee the delivery of timely or useful data}. Even throughput-efficient access schemes can create backlogs, causing flows to age before successful delivery. This has motivated recent work on age-aware random access, namely MisTA~\cite{mista}. In age-threshold schemes, a user attempts access only when the age of its last delivered status exceeds a threshold, thereby thinning the active population and reducing unnecessary contention. However, age alone does not always capture the value of an update. As discussed in Section~\ref{sec:sampling}, the sampling and transmission decision should depend on the application-level loss, which may be non-monotone in age or depend on the semantic relevance of the source state.
	
	\subsection{Goal-Oriented Traffic Shaping}
	
	\begin{figure}[t]
		\centering
		\includegraphics[width=0.76\linewidth]{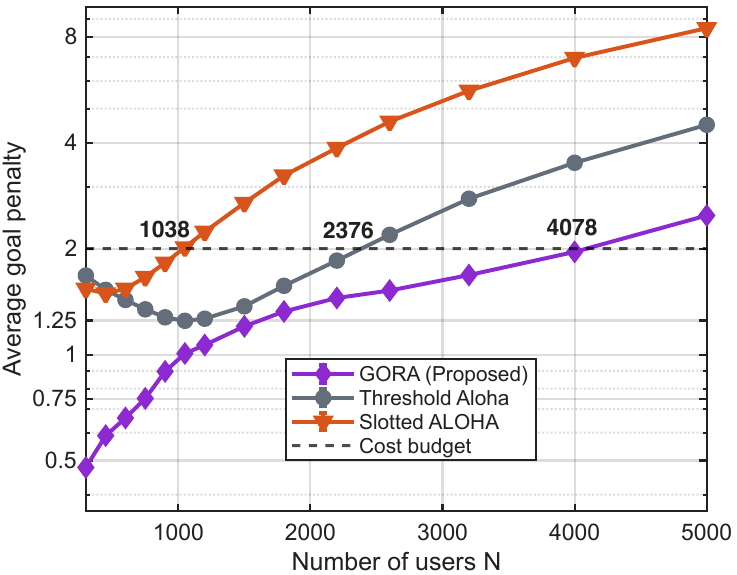}
		\caption{Goal-oriented random access under heterogeneous user goals. Compared with Slotted ALOHA and age-aware Threshold ALOHA, GORA activates users according to delivery-time goal value and supports a substantially larger population under the same normalized goal-penalty budget. At penalty level $2$, it supports approximately $4\times$ as many users as throughput-oriented Slotted ALOHA~\cite{ahsengorajournal}.}
		\label{fig:gora}
	\end{figure}
	
	GORA applies the \textit{value-at-delivery} principle to the access decision. Instead
	of treating packet arrivals as exogenous, each source locally decides
	whether a successful transmission would reduce its goal penalty enough to
	justify an access attempt~\cite{ahsengorajournal}. This performs
	\emph{goal-oriented traffic shaping}: unlike fixed-probability ALOHA, which controls only the offered
	load, and age-threshold ALOHA~\cite{mista}, which thins users by freshness, GORA suppresses sources whose current data carries low marginal task value even when that data is old, and promotes sources with high-value states even when their data is relatively fresh.
	
	Consider a slotted random-access uplink with $N$ low-duty-cycle sources
		sharing a common access opportunity, such as a satellite beam, an orbiter
		relay, or a lander gateway. A packet is successfully delivered if exactly
		one source transmits in a slot; if two or more sources transmit, a collision
		occurs. Each source $i$ belongs to a goal class $k_i$ and has receiver-side
		age $\Delta_i(t)$ before slot $t$. Its application-level loss is represented by
		an age-dependent penalty function $h_{k_i}(\Delta_i(t))$. The goal-oriented
		performance metric is the long-term average goal penalty per source. Each node selects both the access probability and the packet in the buffer to transmit to minimize the long-term average goal penalty per source. 
	
	Fig.~\ref{fig:gora} evaluates a heterogeneous population comprising
		three goal classes (safety-critical, target-age, and intermittent
		monitoring, with proportions $12\%$, $48\%$, and $40\%$) and plots the
		time-average goal penalty per source against the number of users. The relevant performance metric is the largest population each policy
		can sustain under a target penalty budget: a horizontal reading of
		the figure. At a target penalty level of~$2$, Slotted ALOHA sustains
		approximately $1{,}038$ users, Threshold ALOHA $2{,}376$, and GORA
		$4{,}078$, a gain of roughly $4\times$ over throughput-oriented access. The operational advantage of GORA therefore comes
		not from higher raw channel throughput, but from directing access
		attempts toward sources with higher delivery-time relevance, expanding
		the network size that a given task-performance budget can sustain. 
	
	This source-side traffic shaping complements receiver-side techniques such
	as coded random access and successive interference cancellation. Those
	methods improve collision recovery after packets are transmitted, whereas
	GORA reduces low-value contention before packets enter the channel. 
	
	\noindent\fbox{
		\parbox{.95\columnwidth}{\textbf{Takeaway}: GORA allows the MAC layer to benefit from pre-shaped traffic by replacing exogenous data arrivals with \textit{goal-oriented traffic shaping}.} }
	
	\begin{figure}[t]
		\centering
		\includegraphics[width=\columnwidth]{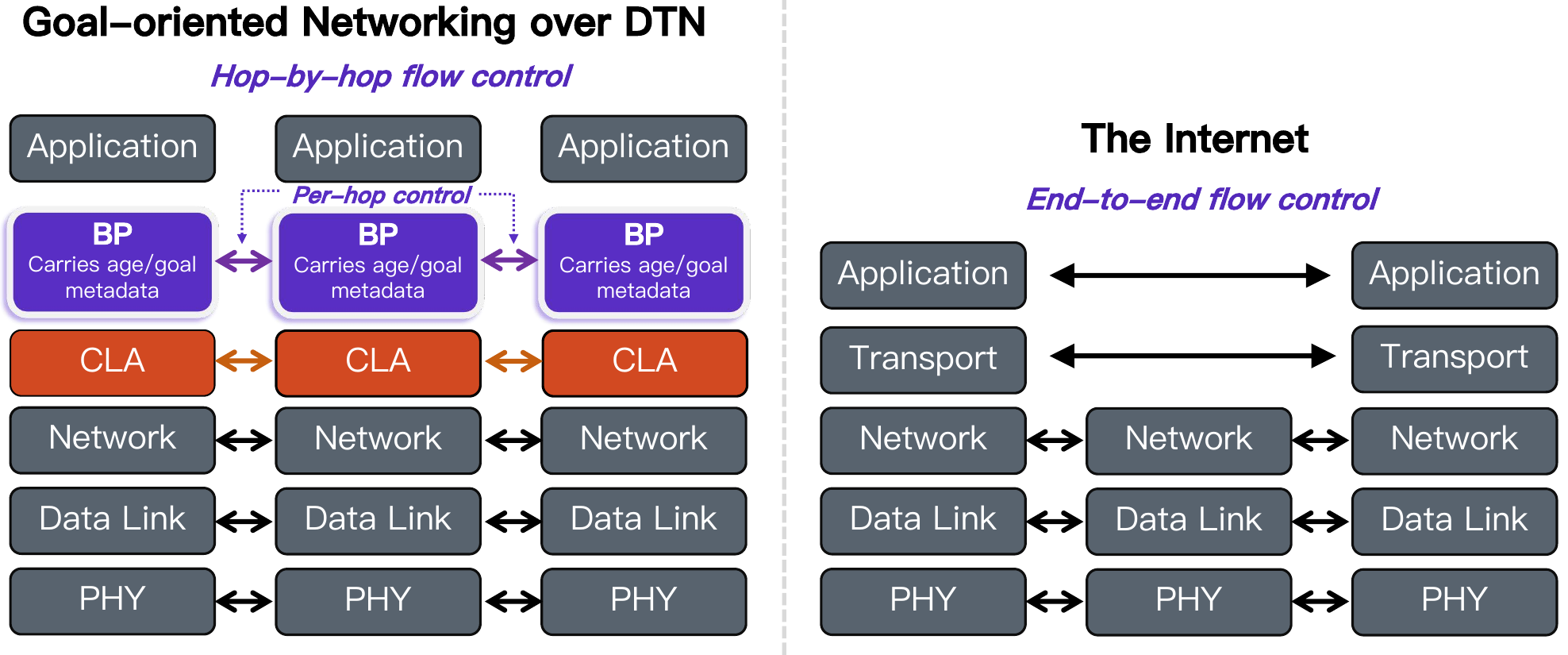}
		\caption{{A contrast of the architectures of DTN and the Internet. The former allows hop-by-hop flow control, whereas in the latter, transport-layer functionalities typically operate end-to-end.}}
		\label{fig:DTNvsInternet}
	\end{figure}
	
	\section{Pillar III: Goal-Oriented Flow Control and Routing}
	Goal-oriented flow control and routing addresses the relay-side
		decisions arising from intermittent contact~(C3) and unpredictable
		traffic volume at contact~(C4). CGR and related DTN mechanisms can identify
	future contacts and next-hop opportunities, but a forwarding relay must
	still decide which route or contact opportunity to use, which buffered bundles deserve the next outgoing contact, which
	bundles should be reprocessed before forwarding, and which bundles should be
	retained or discarded. This is the flow-control and routing side of the \emph{store-compute-forward} regime introduced in
		Section~\ref{sec:spacenets}: a relay stores bundles, computes their
		remaining goal value, optionally reprocesses them in orbit, and routes and forwards
		only those worth the scarce contact capacity. 
	
	Internet-style flow control may be suitable on relatively stable or near-Earth links. However, when long RTTs and disruptions make timely end-to-end feedback unavailable, hop-by-hop decisions become essential. In SpaceNets, the relevant question is how the limited contact volume should be allocated, and over which available routes or contacts.
	
	\subsection{Bundle Protocol as a Goal-Oriented Substrate}
	
	The Bundle Protocol (BP), described in RFC 9171, defines procedures for forwarding data bundles through a DTN. This allows flow control and routing through hop-by-hop decisions, rather than only end-to-end control, in contrast to the architecture of the Internet (see Fig. \ref{fig:DTNvsInternet}). At each hop, a node may have more than one feasible next-hop route or contact, for example a primary and a backup link with different delay and burstiness. The relevant control state is carried by standard protocol metadata rather than by application payloads: the LTP is a common convergence layer for space links, and recently, advanced packet prioritization mechanisms for LTP have been suggested. The recently proposed multicolor LTP protocol allows multiple service classes. Such service classes can be combined with goal-aware
	indicators at the relay layer.
	
	BP is a natural substrate for goal-oriented flow control and routing because it already contains metadata that can support age- and lifetime-aware decisions. At a node running BP, bundle age can be determined from the creation timestamp, or from the Bundle Age extension block when the local clock is not sufficiently accurate. The Lifetime field specifies how long after creation the bundle remains useful; it is especially suitable for goal functions
	whose value decays with age. These fields do not by themselves define a
	goal-oriented policy, but they provide the protocol hooks needed to rank buffered bundles, together with their assigned routes or contacts, according to delivery-time relevance.
	
	\subsection{Goal-Oriented Contact-Start Scheduling}
	The Age and Lifetime fields enable a concrete, goal-aware scheduling step
		when a contact window opens. For each buffered bundle, the Application Agent (AA) or Bundle Protocol Agent (BPA) reads its
		current age and remaining lifetime, then estimates its delivery-time value
		by evaluating the relevant goal function at the age the bundle would reach
		upon arrival, using the contact schedule predicted by CGR. When multiple next-hop routes or contacts are feasible, this
		delivery-time value should guide bundle assignment instead of CGR's default
		minimum-mean-delay rule. A route that is fast on
		average but occasionally very slow may lose more goal value than it saves in
		mean delay~\cite{atasayar2025fresh}. A bundle whose
		raw form would consume more contact capacity than its goal value justifies
		may first be \emph{reprocessed in orbit}, for example compressed, fused with
		nearby observations, or reduced to a task-relevant feature or inference
		result. This is the \emph{compute} step of \textit{store-compute-forward}: it is
		worthwhile when onboard processing is cheaper than the contact capacity it
		saves, or when it increases the goal value carried per transmitted byte.
		Bundles, whether raw or reprocessed, are then ranked by expected goal value
		per unit of contact capacity; the contact volume is filled in rank order;
		and bundles whose Lifetime expires before their earliest feasible delivery
		are dropped. Every step operates on standard BP fields and the existing
	contact plan, with application-level value functions supplied
	by the AA, so age- and lifetime-driven policies can be supported
	within current DTN stacks without requiring a protocol redesign. This makes
	goal-oriented flow control and bundle management deployable within current DTN implementations, an
	attractive property for standardization through CCSDS and IETF.
	\noindent\fbox{\parbox{.95\columnwidth}{\textbf{Takeaway}: DTN relays are more than \textit{passive} buffers. With BP age/lifetime metadata, contact plans, and lightweight task-value annotations, they can decide which bundles deserve a rare contact and which route can deliver them while they still matter, without redesigning the DTN stack.}}

	\section{Interplay of the Three Pillars}
	\label{sec:interplay}
	
	\begin{figure}[t]
		\centering
		\includegraphics[width=0.76\linewidth]{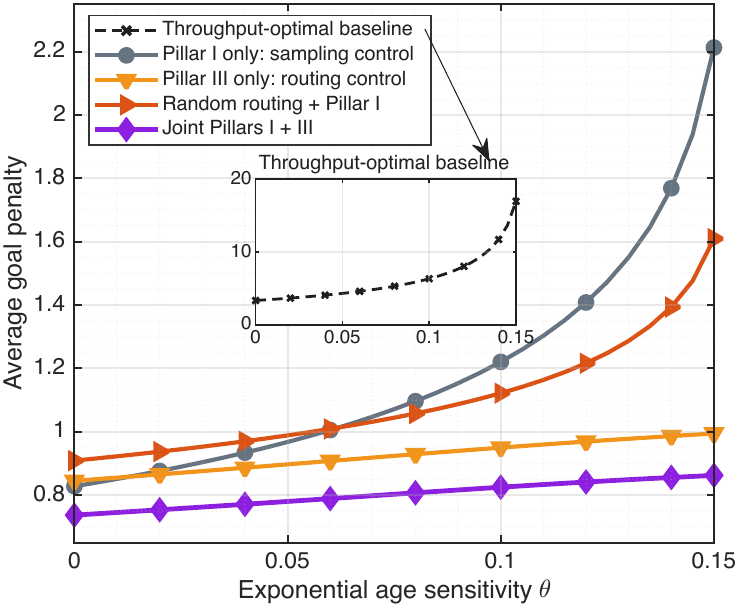}
		\caption{Separated vs. joint goal-oriented control. Either pillar alone recovers most of the
				joint gain; the throughput-optimal (min-delay route, zero-wait sampling) baseline incurs
				nearly $20\times$ the joint policy's penalty at $\theta=0.15$, as rare
				large delays on the fastest route are amplified by the age
				penalty $g_\theta(\Delta)=(e^{\theta\Delta}-1)/\theta$~\cite{atasayar2025fresh}.}
		\label{fig:p1p3}
	\end{figure}
	
	Although each pillar can be optimized individually, in an
		operational SpaceNet they are coupled through a common notion of
		\emph{delivery-time value}. This is not a fast per-packet cross-layer loop:
		long RTTs and intermittent contacts make real-time feedback unrealistic.
		Instead, coordination occurs at contact-cycle time scales through lightweight
		metadata, including age and goal class at the source, queue state and delayed delivery or discard reports at relays.
		
		Fig.~\ref{fig:p1p3} gives a quantitative view of how Pillar~I and Pillar~III interact. Pillar~I controls when updates enter the network; Pillar~III controls which route or contact carries them. When the monitoring task becomes more age-sensitive, this coupling matters: at $\theta=0.15$, the throughput-optimal baseline (min-delay route, zero-wait sampling) incurs nearly $20\times$ the penalty of the joint Pillar~I+III policy. Optimizing either sampling or routing alone removes a large part of the loss, which supports incremental deployment. Yet joint control remains valuable, reducing the average goal penalty by 13\% relative to Pillar~III alone and by 61\% relative to Pillar~I alone.
		
		The same delivery-time-value logic extends along the forward path. \textit{Goal-oriented sampling and scheduling} determines which candidate
		updates are generated, retained, or marked as worth transmission. At the access
		layer, \textit{goal-oriented random access} decides which candidates should
		contend for a scarce contact opportunity, thinning the offered load according
		to goal relevance. \textit{Goal-oriented flow control and routing} then ranks admitted bundles, optionally reprocesses them into
		task-relevant features, and selects the bundles and routes that best preserve
		their value at delivery.
		
		Consider a lunar surface sensor network during a single orbiter pass. Most sensors remain silent because their routine measurements would not affect the event detection; only the few updates with high delivery-time
		value contend for access. The orbiter then ranks admitted bundles, compresses or fuses to raise goal value per transmitted byte,
		and jointly decides which bundles to forward and which route or
		future contact should carry them. After the pass,
		aggregate delivery, expiry, and discard statistics can be piggybacked on later
		beacons or delayed acknowledgements, enabling future sampling, access, and routing
		decisions to be adjusted without real-time feedback. 
	
	\section{Conclusions and Open Challenges} 
	Goal-oriented networking offers a protocol-level perspective for
	SpaceNets in which communication decisions are driven by delivery-time
	value rather than by throughput, delay, or freshness alone. This article
	organized that perspective around three pillars: sampling and
	scheduling decide which information should be generated and served; random
	access decides which sources should contend for scarce contact
	opportunities; routing and flow control introduce goal-oriented forwarding decisions, and in the case of Bundle Protocol, which bundles should be
	reprocessed, forwarded, retained, or discarded at relays. Together, these
	mechanisms implement a \textit{store-compute-forward} view of space networking, in
	which scarce contact capacity is spent on information that is expected to
	remain useful when it reaches the decision point.
	
	The vision presented in this paper opens up avenues for future research on DTN, deep space networks and 6G connectivity. Further work is needed in developing tractable 
	goal oriented KPIs, lightweight policies that align source-side sampling,
	access-layer contention, onboard computation, and relay-side bundle
	management under long delays and intermittent feedback. As SpaceNets evolve toward larger NTN,
	cislunar, and interplanetary deployments, such goal-oriented mechanisms may
	provide a practical path from bit-pipe connectivity to mission-aware
	networking.
	
	\bibliographystyle{IEEEtran}
	\bibliography{GOSPACECommMag}
	
	\begin{IEEEbiographynophoto}{Elif Uysal}
		(Fellow, IEEE) is a Professor of Electrical and Electronics Engineering at the Middle East Technical University (METU), in Ankara, Turkey. She received the Ph.D. degree from Stanford University (2003), S.M. from MIT (1999) and B.S. from METU (1997). She was a recipient of the IEEE INFOCOM 2026 Test of Time Paper Award and the 2025 Selcuk Yasar Award. She received an ERC Advanced Grant in 2024 for her project GO SPACE (Goal Oriented Networking for Space), a TUBITAK BIDEB National Pioneer Researcher Grant (2020), and the Science Academy of Turkey Young Scientist Award (2014). She has chaired the METU Parlar Foundation for Education and Research (2022), is the founder of FRESHDATA Technologies, and is a Fellow of the Artificial Intelligence Industry Alliance (AIIA).
	\end{IEEEbiographynophoto}
	
	\vspace{-2em}
	\begin{IEEEbiographynophoto}{Aimin Li}
		(Member, IEEE) received the B.S. degree (with the
			Best Thesis Award) and the Ph.D. degree (nominated for the Best
			Dissertation Award) from Harbin Institute of Technology (Shenzhen),
			China. He was a visiting researcher with the Institute for Infocomm
			Research (I2R), A*STAR, Singapore, from 2023 to 2024. He is
			currently a postdoctoral researcher with METU, Ankara, Turkey.
	\end{IEEEbiographynophoto}
	
\end{document}